\documentclass[conference]{IEEEtran}
\IEEEoverridecommandlockouts
\usepackage{cite}
\usepackage{amsmath,amssymb,amsfonts}
\usepackage{algorithmic}
\usepackage{graphicx}
\usepackage{textcomp}
\usepackage{xcolor}

\usepackage{tikz}
\usepackage{pgfplots}
\usepackage{amsmath}
\usepackage{subcaption}
\usepackage{siunitx}
\usepackage{adjustbox}
\usepackage{multirow}

\usetikzlibrary{
    calc,
    pgfplots.groupplots,
    patterns
}
\pgfplotsset{
    compat=1.3,
}

\def\BibTeX{{\rm B\kern-.05em{\sc i\kern-.025em b}\kern-.08em
    T\kern-.1667em\lower.7ex\hbox{E}\kern-.125emX}}

\DeclareMathOperator*{\argmax}{arg\,max}

\begin{document}

\title{A Novel Signal Processing Strategy for Short-Range Laser Feedback Interferometry Sensors
\thanks{The project is supported by the Chips Joint Undertaking (Chips JU) and its members, including top-up funding by Denmark, Germany, Netherlands, Sweden, under grant agreement No. 101139942.}
}

\author{\IEEEauthorblockN{Alexander Zimmer}
\IEEEauthorblockA{\textit{Human-Centered Tech. for Learning} \\
\textit{Technical University of Munich}\\
Munich, Germany \\
0009-0003-8800-104X}
\and
\IEEEauthorblockN{Johannes Meyer}
\IEEEauthorblockA{\textit{Corporate Sector Research and Advance Eng.} \\
\textit{Robert Bosch GmbH}\\
Renningen, Germany \\
0000-0001-8370-2603}
\and
\IEEEauthorblockN{Enkelejda Kasneci}
\IEEEauthorblockA{\textit{Human-Centered Tech. for Learning} \\
\textit{Technical University of Munich}\\
Munich, Germany \\
0000-0003-3146-4484}
}

\IEEEoverridecommandlockouts

\maketitle

\IEEEpubidadjcol

\begin{abstract}
The rapid evolution of wearable technologies, such as AR glasses, demands compact, energy-efficient sensors capable of high-precision measurements in dynamic environments. Traditional Frequency-Modulated Continuous Wave (FMCW) Laser Feedback Interferometry (LFI) sensors, while promising, falter in applications that feature small distances, high velocities, shallow modulation, and low-power constraints. We propose a novel sensor-processing pipeline that reliably extracts distance and velocity measurements at distances as low as 1 cm. As a core contribution, we introduce a four-ramp modulation scheme that resolves persistent ambiguities in beat frequency signs and overcomes spectral blind regions caused by hardware limitations. Based on measurements of the implemented pipeline, a noise model is defined to evaluate its performance and sensitivity to several algorithmic and working point parameters. We show that the pipeline generally achieves robust and low-noise measurements using state-of-the-art hardware.
\end{abstract}

\begin{IEEEkeywords}
Laser Sensor, LFI, SMI, FMCW, VCSEL, Low-power signal processing
\end{IEEEkeywords}

\date{March 10, 2025}

\section{Introduction}

Emerging applications like gaze-based interaction\cite{meyergazegesture} and continuous focus tracking demand eye sensors that combine sub-milliwatt power, millimeter-scale precision, ambient-light robustness, and imperceptible form factors, a combination unachievable with camera-based systems. While static setups based on Laser Feedback Interferometry (LFI) sensors  theoretically address these needs\cite{capelli,meyerstaticlfi,meyerambilight}, inherent limitations hinder their practical adoption in short-range, high-velocity regimes. Existing Frequency-modulated continuous wave (FMCW) LFI methods sacrifice minimum working distance and modulation capabilities or fail to disambiguate high-velocity measurements, thus limiting wearable integration.

To bridge this gap, we introduce a signal processing pipeline that robustly extracts distance and velocity measurements from an LFI sensor, optimized for stringent power, size, and dynamic-range constraints

The system leverages an ultra-compact vertical-cavity surface-emitting laser (VCSEL) with a wavelength of \qty{848}{\nano\metre} and a collimating optical element, achieving a package diameter smaller than \qty{2}{\milli\metre}, thus ensuring unobtrusive integration in smart glasses.

Our key contributions are:

\begin{itemize}
    \item A full signal processing pipeline for FMCW LFI sensors.
    \item A novel modulation scheme and processing algorithm to support the combination of small distances, high velocities, and shallow modulation ramps.
    \item A noise model to evaluate the pipeline and choose the algorithm and working point parameters.
\end{itemize}

Possible application areas include highly integrated wearable eye-tracking and gaze-gesture sensors in smart glasses, allowing for novel approaches in user interaction, healthcare, or behavior research even in consumer devices.

\section{Related work}

\subsection{Foundations of LFI}

Self-mixing interferometry (SMI) in Lasers, first formalized by Lang and Kobayashi\cite{langkobayashi}, enables compact velocimetry and vibrometry sensors\cite{smitutorial}. Small displacements can be measured by directly analyzing the waveform as described by Donati et al.\cite{donati}. To get absolute distance and velocity measurements, the application of the FMCW technique to LFI sensors was introduced by Beheim et al.\cite{Beheim:86}, initially via counting mode hops. Gouaux et al.\cite{Gouaux:98} proposed using the beat frequency calculated from this count. More recent applications instead utilize a Fast Fourier Transform (FFT) and peak detection to get the beat frequency, as introduced by Tucker et al.\cite{tucker}, which is still state of the art\cite{trumpf2022}.

For two known modulation ramp slopes $S = \frac{\mathrm{d}f}{\mathrm{d}t}$ and an average emitted frequency $f_e$, the sensor measurements can be determined from the beat frequencies $f_b$ of both ramps. The distance measurement $R$ is given by

\begin{equation}
\label{eq:distance}
    R = \frac{c \left(\pm f_{b,1} - \pm f_{b,2}\right)}{2 (S_1 - S_2)} \; , \; S_1 \neq S_2
\end{equation}

And the velocity $v$ by

\begin{equation}
\label{eq:velocity}
    v = \frac{c \left(\pm f_{b,2} S_1 - \pm f_{b,1} S_2\right)}{f_e \left(S_1 - S_2\right)} \; , \; S_1 \neq S_2 \; , \; f_e \neq 0
\end{equation}

Equations \ref{eq:distance} and \ref{eq:velocity} are derived directly from theory.

\subsection{Limitations in short-range regimes}

In literature, distance- and velocity-related frequencies are typically provided as follows\cite{trumpf2008,meyerstaticlfi,lidar}:

\begin{equation}
\label{eq:distance_simp}
    f_R = \frac{f_{\mathrm{up}} + f_{\mathrm{down}}}{2}
\end{equation}

\begin{equation}
\label{eq:velocity_simp}
    f_v = \frac{f_{\mathrm{up}} - f_{\mathrm{down}}}{2}
\end{equation}

Equations \ref{eq:distance_simp} and \ref{eq:velocity_simp} are equivalent to Equations \ref{eq:distance} and \ref{eq:velocity}, respectively, for symmetric triangular modulation patterns under the assumption that the frequency shift induced by the distance dominates over that by the velocity, even though this constraint is typically omitted in the existing literature. The sign of the beat frequency is generally ambiguous, and these simplified equations are not universally valid.

The wearable eye-tracking use-case is unique in that it has low expected distance measurements (between \qty{2}{\centi\metre} and \qty{5}{\centi\metre}) and a wide range of surface velocities up to approximately \qty{0.1}{\metre\per\second}, resulting from a maximum rotational speed of the eye of \qty{500}{\degree\per\second}\cite{meyerstaticlfi}. In addition, modulation ramp slopes are constrained by the hardware. Therefore, both cases for each frequency sign are possible, so Equations \ref{eq:distance} and \ref{eq:velocity} provide no definite solution.

Furthermore, due to the self-mixing nature of the sensor, a hardware high-pass filter is applied to the signal from the diode to suppress the low-frequency component of the modulation signal, which was already proposed by Tucker et al.\cite{tucker}. Thus, low beat frequencies cannot be reliably measured. This leads to the issue of blind regions in the distance/velocity measurement space where the beat frequencies of one or more ramps are out of range, resulting in systematically false measurements.

\subsection{Critical advancements}

To our knowledge, these issues have not yet been described in the literature. This work focuses on developing a sensor processing approach that works in short-range regimes by solving the issues regarding sign ambiguities and blind regions.

\section{LFI sensor processing pipeline}

\begin{figure}
    \centering
    \includegraphics[width=\linewidth]{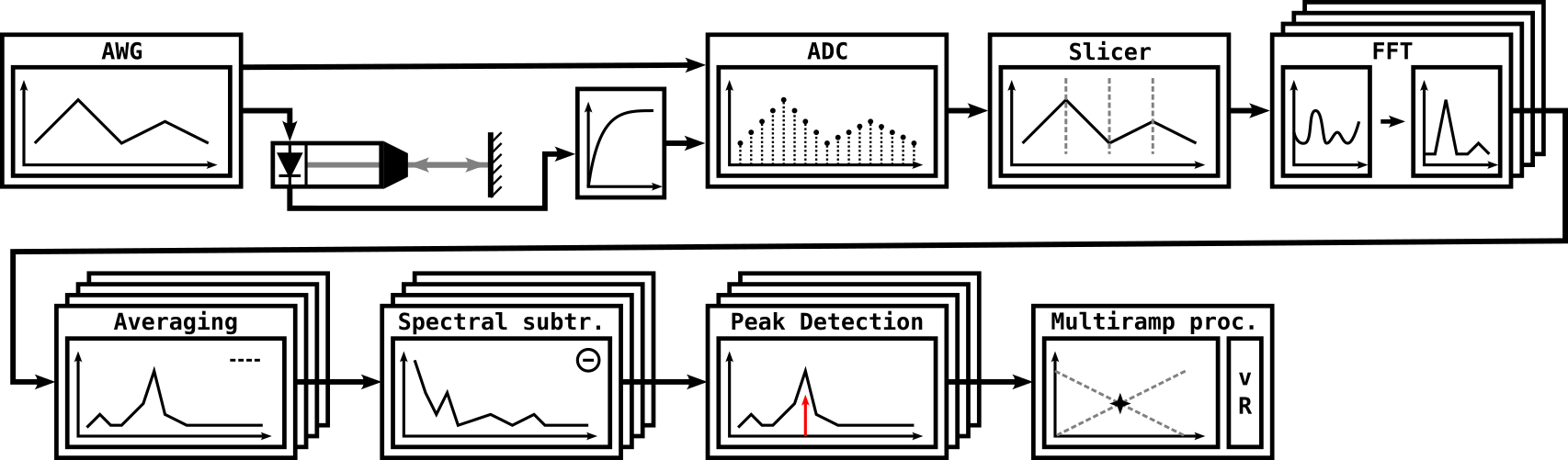}
    \caption[]{Block diagram of the proposed sensor processing pipeline. The modulation signal generated by an arbitrary waveform generator (AWG) is channeled into the Sensor and, in sync with the sensor output, recorded by an analog-to-digital converter (ADC) unit. It is then sliced into individual modulation segments and transformed into the frequency domain. A peak detection algorithm is applied after averaging multiple modulation pattern cycles and filtering the noise floor. The extracted peak frequencies are then processed to return the final measurement values $v, R$ for each time sample.}
    \label{fig:dspPipeline}
\end{figure}

We introduce a novel sensor processing pipeline based on existing FMCW LFI algorithms\cite{tucker}\cite{trumpf2022}, defined by a working point including an extended modulation pattern and a digital signal processing algorithm responsible for extracting the distance and velocity from the raw time series measurements. The pipeline is visualized in Figure \ref{fig:dspPipeline}.

\subsection{Modulation scheme}

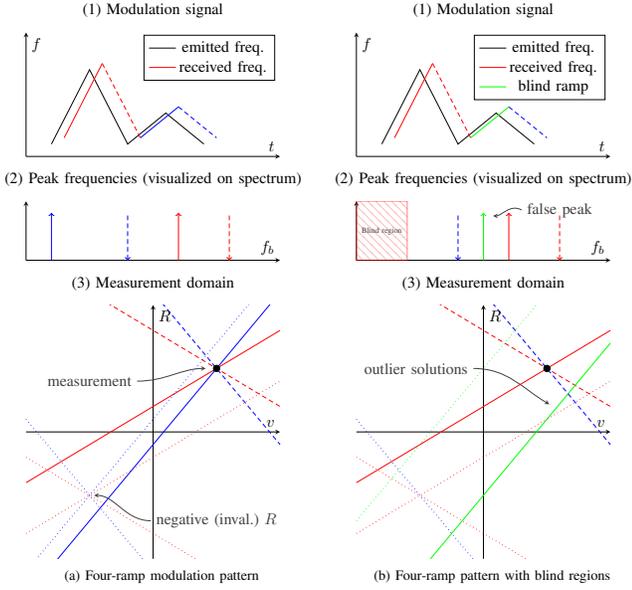
\begin{figure}
\centering
\resizebox{\linewidth}{!}{
\begin{subfigure}{.4\textwidth}
\begin{tikzpicture}
\begin{groupplot}[
    group style={
        group size=1 by 3,
    }
]

\nextgroupplot[
    title={(1) Modulation signal},
    view={0}{90},
    axis lines=center,
    ticks=none,
    xmin=0,xmax=1,ymin=0,ymax=1,
    xlabel={$t$},ylabel={$f$},
    xticklabel=\empty,yticklabel=\empty,
    width=\textwidth,height=.6\textwidth
]

\addplot [no marks] coordinates {(.1,.1) (.25,.7) (.4,.1) (.55,.35) (.7,.1)};
\addlegendentry{emitted freq.}

\addplot [red, no marks] coordinates {(.15,.15) (.3,.75)};
\addlegendentry{received freq.}
\addplot [red, no marks, densely dashed] coordinates {(.3,.75) (.45,.15)};
\addplot [blue, no marks] coordinates {(.45,.15) (.6,.4)};
\addplot [blue, no marks, densely dashed] coordinates {(.6,.4) (.75,.15)};

\nextgroupplot[
    title={(2) Peak frequencies (visualized on spectrum)},
    view={0}{90},
    axis lines=center,
    ticks=none,
    xmin=0,xmax=1,ymin=0,ymax=1,
    xlabel={$f_b$},
    xticklabel=\empty,yticklabel=\empty,
    width=\textwidth,height=.4\textwidth
]

\addplot [red, ->, no marks] coordinates {(.6,0) (.6,.8)};
\addplot [red, <-, no marks, densely dashed] coordinates {(.8,0) (.8,.8)};
\addplot [blue, ->, no marks] coordinates {(.1,0) (.1,.8)};
\addplot [blue, <-, no marks, densely dashed] coordinates {(.4,0) (.4,.8)};

\nextgroupplot[
    title={(3) Measurement domain},
    view={0}{90},
    axis lines=center,
    ticks=none,
    xmin=-1,xmax=1,ymin=-1,ymax=1,
    xlabel={$v$},ylabel={$R$},
    xticklabel=\empty,yticklabel=\empty,
    width=\textwidth,height=\textwidth
]

\addplot [black, only marks] coordinates {(.5,.5)};

\addplot [red, no marks] coordinates {(-1,-.4) (1,.8)};
\addplot [red, no marks, densely dashed] coordinates {(-1,1.4) (1,.2)};
\addplot [blue, no marks] coordinates {(-1,-1.3) (1,1.1)};
\addplot [blue, no marks, densely dashed] coordinates {(-1,2.3) (1,-.1)};

\addplot [red, no marks, dotted] coordinates {(-1,-.8) (1,.4)};
\addplot [red, no marks, dotted] coordinates {(-1,-.2) (1,-1.4)};
\addplot [blue, no marks, dotted] coordinates {(-1,-1.1) (1,1.3)};
\addplot [blue, no marks, dotted] coordinates {(-1,.1) (1,-2.3)};

\node [darkgray] (origin) at (axis cs:.5,-.7) {negative (inval.) $R$};
\draw [darkgray, shorten <=.1cm, stealth-] (axis cs:-.5,-.5) to [out=0,in=180] (origin.west);

\node [darkgray] (origin) at (axis cs:-.5,.4) {measurement};
\draw [darkgray, shorten <=.1cm, stealth-] (axis cs:.45,.5) to [out=180,in=0] (origin.east);

\end{groupplot}
\end{tikzpicture}
\caption{Four-ramp modulation pattern}
\end{subfigure}
\begin{subfigure}{.4\textwidth}
\begin{tikzpicture}
\begin{groupplot}[
    group style={
        group size=1 by 3,
    }
]

\nextgroupplot[
    title={(1) Modulation signal},
    view={0}{90},
    axis lines=center,
    ticks=none,
    xmin=0,xmax=1,ymin=0,ymax=1,
    xlabel={$t$},ylabel={$f$},
    xticklabel=\empty,yticklabel=\empty,
    width=\textwidth,height=.6\textwidth
]

\addplot [no marks] coordinates {(.1,.1) (.25,.7) (.4,.1) (.55,.35) (.7,.1)};
\addlegendentry{emitted freq.}

\addplot [red, no marks] coordinates {(.15,.15) (.3,.75)};
\addlegendentry{received freq.}
\addplot [green, no marks] coordinates {(.45,.15) (.6,.4)};
\addlegendentry{blind ramp}
\addplot [red, no marks, densely dashed] coordinates {(.3,.75) (.45,.15)};
\addplot [blue, no marks, densely dashed] coordinates {(.6,.4) (.75,.15)};

\nextgroupplot[
    title={(2) Peak frequencies (visualized on spectrum)},
    view={0}{90},
    axis lines=center,
    ticks=none,
    xmin=0,xmax=1,ymin=0,ymax=1,
    xlabel={$f_b$},
    xticklabel=\empty,yticklabel=\empty,
    width=\textwidth,height=.4\textwidth
]

\draw[red, pattern=north west lines, pattern color=red, opacity=0.5] (axis cs:0,0) rectangle (axis cs:.2,1);

\addplot [red, ->, no marks] coordinates {(.6,0) (.6,.8)};
\addplot [red, <-, no marks, densely dashed] coordinates {(.8,0) (.8,.8)};
\addplot [green, ->, no marks] coordinates {(.5,0) (.5,.8)};
\addplot [blue, <-, no marks, densely dashed] coordinates {(.4,0) (.4,.8)};

\node [darkgray] (origin) at (axis cs:.8,.85) {false peak};
\draw [darkgray, shorten <=.1cm, stealth-] (axis cs:.52,.75) to [out=0,in=150] (origin.west);

\node [darkgray] (origin) at (axis cs:.1,.5) {\tiny {Blind region}};

\nextgroupplot[
    title={(3) Measurement domain},
    view={0}{90},
    axis lines=center,
    ticks=none,
    xmin=-1,xmax=1,ymin=-1,ymax=1,
    xlabel={$v$},ylabel={$R$},
    xticklabel=\empty,yticklabel=\empty,
    width=\textwidth,height=\textwidth
]

\addplot [black, only marks] coordinates {(.5,.5)};

\addplot [red, no marks] coordinates {(-1,-.4) (1,.8)};
\addplot [red, no marks, densely dashed] coordinates {(-1,1.4) (1,.2)};
\addplot [green, no marks] coordinates {(-1,-1.7) (1,.7)};
\addplot [blue, no marks, densely dashed] coordinates {(-1,2.3) (1,-.1)};

\addplot [red, no marks, dotted] coordinates {(-1,-.8) (1,.4)};
\addplot [red, no marks, dotted] coordinates {(-1,-.2) (1,-1.4)};
\addplot [green, no marks, dotted] coordinates {(-1,-.7) (1,1.7)};
\addplot [blue, no marks, dotted] coordinates {(-1,.1) (1,-2.3)};

\node [darkgray] (origin) at (axis cs:-.53,.5) {outlier solutions};
\draw [darkgray, shorten <=.1cm, stealth-] (axis cs:.55,.2) to [out=150,in=0] (origin.east);

\end{groupplot}
\end{tikzpicture}
\caption{Four-ramp pattern with blind regions}
\end{subfigure}
}
\caption[]{Visualization of the emitted and received light (1), the detected peak frequencies of each ramp as a visualization on the frequency spectrum (2), and the measurement space with two mirrored solution lines for each ramp (3), for one complete modulation cycle with (b) and without (a) considering blind regions. The four-ramp modulation pattern resolves issues with blind spots and sign ambiguities. Even if one ramp is an outlier, there is still a definite solution determined by the remaining three ramps.}
\label{fig:fourramps}
\end{figure}

To solve the issues of sign ambiguity and blind regions, we propose an extended four-ramp modulation pattern with two differently sloped triangles. Adding a third ramp allows determining the signs of all beat frequencies, while the fourth ramp adds redundancy if one ramp becomes blind. Figure \ref{fig:fourramps} explains this modulation scheme.

\begin{figure}
    \centering
    
    \begin{subfigure}{\linewidth}
    \includegraphics[width=\linewidth]{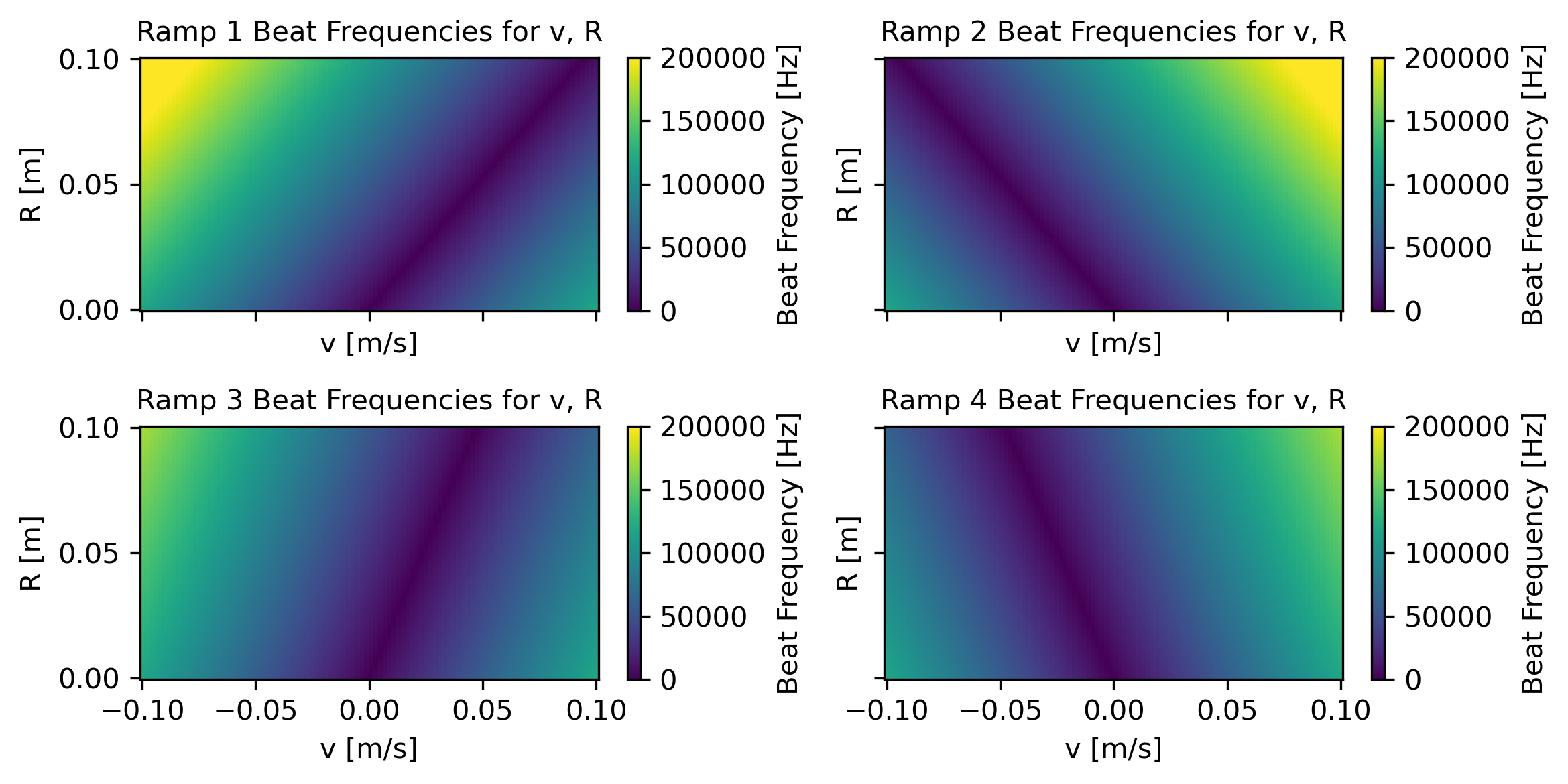}
    \caption{Expected beat frequencies for each of the four modulation ramps in the $v, R$ measurement space.}
    \end{subfigure}

    \begin{subfigure}{\linewidth}
    \centering
    \includegraphics[width=.5\linewidth]{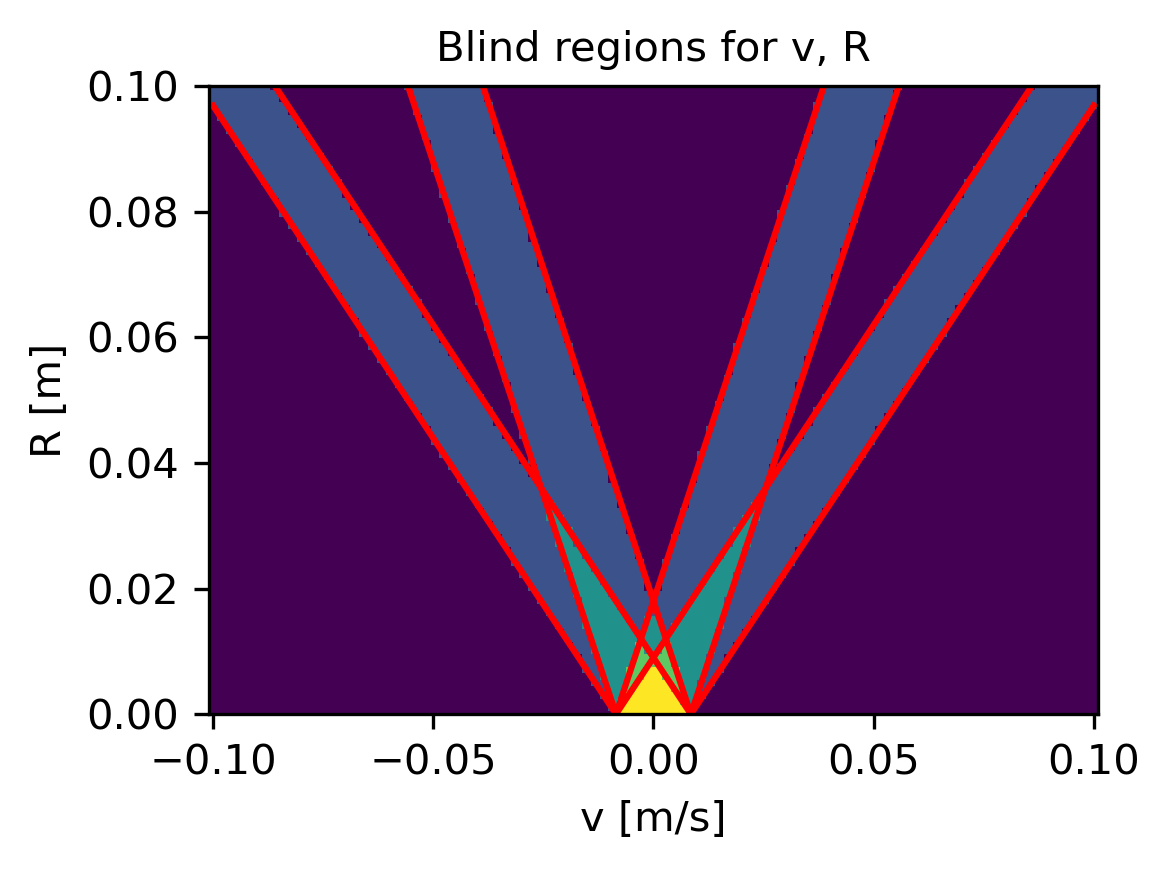}
    \caption{As beat frequencies below $10\,\mathrm{kHz}$ cannot be measured, there are blind regions in the measurement space (shaded areas). With a four-ramp modulation scheme, an unambiguous measurement is possible except for the areas where multiple blind regions overlap (green/yellow areas). Those constitute the minimum reliable distance.}
    \end{subfigure}
    
    \caption[]{Blind regions}
    \label{fig:blindspots}
\end{figure}

Unless both $v$ and $R$ are small, only one peak can be in a blind region simultaneously. The minimum distance at which this is still valid, regardless of velocity, can be determined from the sensor parameters based on the intersections of blind regions in the measurement space. The blind region issue is visualized in Figure \ref{fig:blindspots}.

The key parameters of the modulation scheme are the ramp duration $f_\mathrm{ramp}$, the slope of the steepest ramp $S = \frac{\mathrm{d}f}{\mathrm{d}t}$ and the ratio $\mathrm{rt}$ of the smaller to the steeper triangle. The average emitted frequency $f_e = \frac{c}{\lambda_e} = \frac{c}{848\,\mathrm{nm}}$ depends on the hardware.

\subsection{Sensor processing}

After data acquisition, the signal measured from the diode is synchronized with the modulation pattern and sliced into individual ramps. Those segments are zero-padded, and a Hamming window is applied to smooth the signals. They are then transformed into the frequency domain via an FFT. The resulting spectra $X(k)$ are averaged over several modulation cycles to reduce noise.

\begin{equation}
    \hat{X}_t(k) = \frac{1}{n_{\mathrm{avg}}} \sum_{i = t - n_{\mathrm{avg}}}^{t} X_i(k)
\end{equation}

The spectra are denoised via adaptive spectral subtraction using a calibrated, heavily averaged reference spectrum $\hat{D}(k)$ without a measurement target multiplied by an over-subtraction factor $\alpha$. The noise of the individual frequency bins $\hat{\Sigma}(k)$ is also estimated from the calibration measurement and is added to the spectral subtraction with a factor $\beta$. Spectral flooring is applied to avoid mathematical issues with the peak detection algorithm.

\begin{equation}
    \hat{X}(k) = \max \left(X(k) - \alpha \hat{D}(k)  - \beta \hat{\Sigma}(k), \mathbf{0}\right)
\end{equation}

A two-step peak detection and bin-interpolation algorithm is applied to reliably extract the beat frequencies from the filtered spectra. This approach avoids the sensor resolution being limited to the FFT bin count and reduces noise, as spectral peaks typically span more than one bin and are Gaussian-shaped. First, the bin with the highest intensity is selected. Then, a Gaussian function is fitted to the window of surrounding frequency bins using a least-squares algorithm, and the beat frequency is estimated.

\begin{equation}
\label{eq:gaussinterpol}
    \hat{f_b} = \argmax_f \; \hat{a} \exp \left( -\frac{\left(f - \hat{b}\right)^2}{2\hat{c}^2} \right) = \hat{b}
\end{equation}

Alternatively, the weighted average of that window (with frequencies $F(k)$) can be used.

\begin{equation}
    \hat{f_b} = \frac{1}{n_{\mathrm{bins}}} \sum_{k = k_{\mathrm{max}} - n_{\mathrm{bins}}/2}^{k_{\mathrm{max}} + n_{\mathrm{bins}}/2} \hat{X}(k) F(k)
\end{equation}

The distance and velocity measurements are calculated from the estimated beat frequencies via Equations \ref{eq:distance} and \ref{eq:velocity}. Using four ramps leads to an over-determined system. A low peak intensity indicates an unreliable measurement, typically caused by blind regions, so only the three ramps with the highest intensity are considered. Measurements for all combinations of signs for the beat frequency and ramp pairs are calculated. The two inverse combinations of signs for which all calculated measurements from the possible ramp pair combinations are clustered the best are selected, and the respective means are taken. Only one of these solutions has a positive distance measurement and is correct. %The approach is visualized in Figure \ref{fig:combined_argmax}.

The key parameters of the sensor processing pipeline are the window function, the FFT bin count, the amount of spectral averaging $n_\mathrm{avg}$, the over-subtraction factors $\alpha, \beta$, and the bin interpolation algorithm and window size.
%TODO hieraus stichpunkte?

\section{Evaluation}

The proposed sensor processing pipeline was evaluated regarding resolution, noise, and minimum reliable distance.

\subsection{Resolution}

Due to bin interpolation, the achievable resolution of the sensor is theoretically infinite, even for a small FFT bin count.

The sample rate, defined by the duration of a complete modulation pattern cycle, was set to $1\,\mathrm{kHz}$ without averaging and can be further increased by adapting hardware filters.

\subsection{Noise}

\begin{figure}
    \centering
    \includegraphics[width=\linewidth]{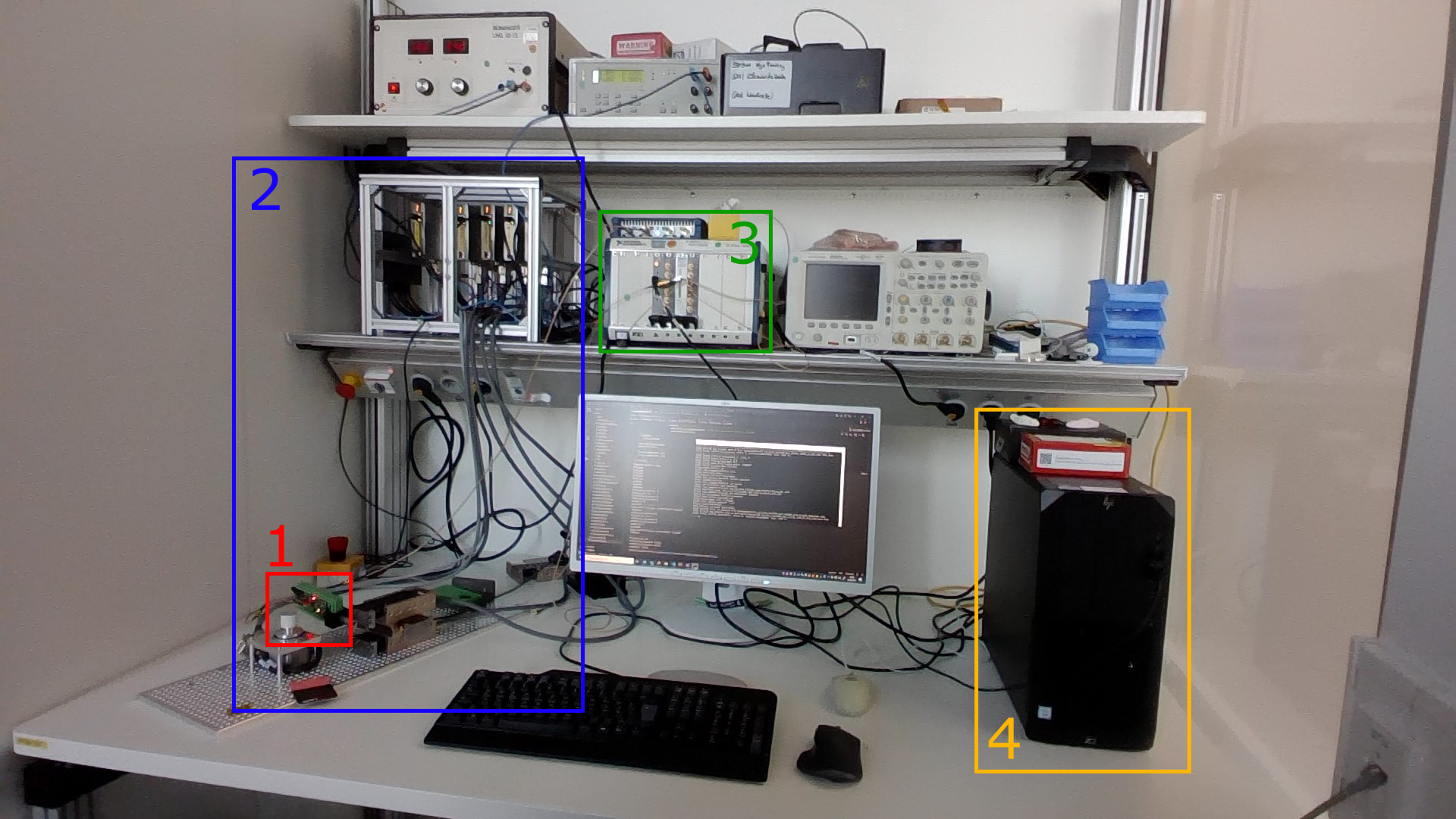}
    \caption[]{(a) Experimental lab setup consisting of the sensor hardware (1), a servo-based setup with a rotating cylinder as a measurement target and a stack of linear axes that allow the automatic adjustment of the sensor position (2), AWG and ADC units (3), and a computer for data processing (4).}
    \label{fig:expsetup}
\end{figure}

To evaluate and characterize the pipeline and choose optimal modulation and sensor processing parameters, a noise model is defined to estimate the relation of the Gaussian beat frequency noise to the parameters and measurement values. Those relationships were shown to be approximately log-log-linear. Thus, the model is defined accordingly:

\begin{equation}
\label{eq:noisemodel}
\begin{split}
    \log\left(\sqrt{n_{\mathrm{avg}}} \sigma_{f_b}\right) = &a_1 \log\left(f_{\mathrm{ramp}}\right) + a_2 \log\left(S\right) \\
    &+ a_3 \log\left(f_b\right) + a_4 \log\left(v\right) + a_5 \log\left(R\right) + b
\end{split}
\end{equation}

The modulation pattern's peak values were fixed, so $f_\mathrm{ramp}$ determines the slope $S$. The parameter $\mathrm{rt}$ is primarily relevant for the minimum reliable distance and omitted from the model.

Sensor processing parameters, except for $n_\mathrm{avg}$, were fixed to an FFT bin count of 2048, an interpolation window size of 25, and over-subtraction factors $\alpha = 1, \beta = 0$, values which performed well in earlier samples. Experimentally, the weighted average approach was shown to slightly outperform the Gaussian interpolation while being more robust and less computationally expensive. Averaging reduces measurement noise at the cost of a lower sample rate proportional to $\sqrt{n_\mathrm{avg}}$. As the noise is Gaussian, this aligns with the theory. %D was calibrated

The noise of the measurements $\sigma_R$ and $\sigma_v$ can be estimated from the beat frequency noise of two ramps $\sigma_{f_{b,n}}$.

\begin{equation}
\sigma_R = \frac{c \left( \sqrt{\sigma_{f_{b,1}}^2 + \sigma_{f_{b,2}}^2}\right)}{2 \left(S_1 - S_2\right)}
\end{equation}

\begin{equation}
\sigma_v = \lambda_e \frac{\sqrt{S_1^2 \sigma_{f_{b,2}}^2 + S_2^2 \sigma_{f_{b,1}}^2}}{S_1 - S_2}
\end{equation}

A prototype of the sensor hardware was mounted on linear axes and aimed at a stepper-motor-driven cylinder. The setup is shown in Figure \ref{fig:expsetup}. 1400 measurements were recorded along a grid of working points and measurement configurations, each lasting \qty{2.5}{\second}. The pipeline was executed several times with different configurations. The model defined in Equation \ref{eq:noisemodel} was then fit to the measured noise data.

\begin{figure}
    \centering
    \includegraphics[width=.95\linewidth]{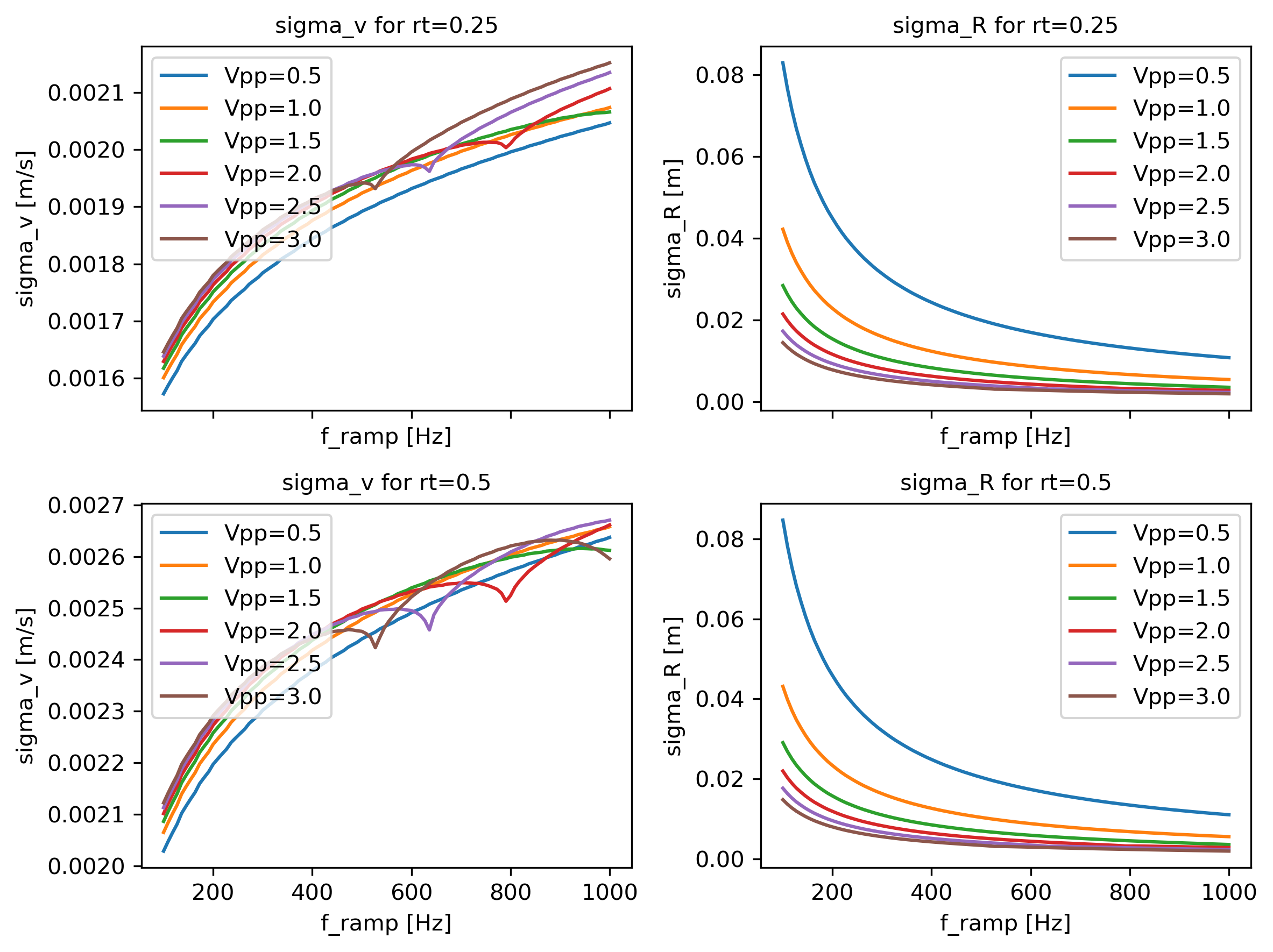}
    \caption[]{Estimated average noise levels $\sigma_v$ and $\sigma_R$ for different working points and without averaging according to the noise model. Higher values of $f_\mathrm{ramp}$ and $\mathrm{V_{pp}}$ correspond to steeper working points. Generally, $\sigma_v$ increases for steeper working points, while $\sigma_R$ decreases. Non-linearities in $\sigma_v$ and $\sigma_R$ are due to the simultaneous influence of multiple $\sigma_{f_b}$. A working point with $\mathrm{rt} = 0.5$ is slightly noisier than with $\mathrm{rt} = 0.25$.} %TODO Vpp erklären oder plot neu machen zefix
    \label{fig:pred_noise_wps}
\end{figure}

Figure \ref{fig:pred_noise_wps} visualizes the estimated effects of the working point parameters on the noise. In the best case without averaging, $\sigma_v$ lies around \qty{1}{\milli\metre\per\second}, while $\sigma_R$ is around \qty{1}{\milli\metre}. Higher velocities and distances result in noisier measurements.

Contrary to theory, where the velocity noise should be decoupled from the slope, $\sigma_v$ tends to increase for steeper slopes. A non-optimal wavenumber linearization might explain this. At the same time, $\sigma_R$ strongly decreases, which is expected due to the better utilization of the FFT bin range.

 %TODO hier bei allen plots nochmal die Texte überarbeiten...

\subsection{Minimum reliable distance}

The optimal $\mathrm{rt}$ for optimizing the minimum distance lies between 0.25 and 0.5, and generally, steeper working points are preferable. With a hardware threshold frequency of $10\,\mathrm{kHz}$, distance measurements above \qty{1}{\centi\metre} are reliably possible.

\section{Conclusion}

This work bridges a fundamental functionality gap in existing LFI sensors. By integrating a novel four-ramp modulation pattern with established LFI sensor processing algorithms, we resolve the issues of measurement ambiguities and blind regions for low-distance, high-velocity, and shallow modulation regimes. The approach is generally suitable for power-constrained systems aiming to achieve low computational complexity, as all algorithms are efficient to implement.

The evaluation on a hardware setup proves the applicability of the pipeline. The proposed noise model can be used as a baseline to optimize parameters and solve trade-offs.

While resolving fundamental issues in existing algorithms, the achieved sensor noise is still about one order of magnitude larger than comparable systems using different hardware\cite{meyerstaticlfi}. So far, the extended modulation scheme is mainly relevant for cases where designing the hardware for steeper ramps to push the blind regions out of the relevant measurement space is infeasible and for systems using more complex modulation schemes, most notably multi-target-capability known from Radar\cite{radarmulti} requiring substantial variations in ramp slopes. More work is necessary to reduce noise along the pipeline.

Overall, the proposed pipeline is robust and can be used to choose working points, improve the hardware, and as a basis for implementing an optimized, product-ready algorithm for novel, constrained use cases in wearables and beyond.

%\section*{Acknowledgment}

\bibliography{literature}
\bibliographystyle{plain}

\end{document}